# P1.81 AN INITIAL ASSESSMENT OF THE GOES MICROBURST WINDSPEED POTENTIAL INDEX

Kenneth L. Pryor
Center for Satellite Applications and Research (NOAA/NESDIS)
Camp Springs, MD

## 1. INTRODUCTION

A suite of products has been developed and evaluated to assess hazards presented by convective downbursts (Fujita 1985) to aircraft in flight derived from the current generation of Geostationary Operational Environmental Satellite (GOES 8-P). The existing suite of GOES-sounder (Menzel et al. 1998) derived microburst products are designed to accurately diagnose risk based on conceptual models of favorable environmental profiles. Pryor and Ellrod (2004) and Pryor and Ellrod (2005) outlined the development of a Geostationary Operational Environmental Satellite (GOES) sounder-derived wet microburst severity index (WMSI) product to assess the potential magnitude of convective downbursts over the eastern United States, incorporating convective available potential energy (CAPE) as well as the vertical theta-e difference (TeD) (Atkins and Wakimoto 1991) between the surface and mid-troposphere. In addition, Pryor (2006a) developed a GOES Hybrid Microburst Index (HMI) product intended to supplement the GOES WMSI product over the United States Great Plains region. The HMI product infers the presence of a convective boundary layer (CBL) (Stull 1988, Sorbjan 1989) by incorporating the sub-cloud temperature lapse rate as well as the dew point depression difference between the typical level of a warm season convective cloud base over the Great Plains and the sub-cloud layer. Thus, the WMSI and HMI algorithms are designed to parameterize updraft and downdraft instability within a convective storm. In addition, the HMI algorithm describes the moisture stratification of the sub-cloud layer that may result in further downdraft acceleration due to evaporative cooling. The intense downdraft subsequently produces strong and potentially damaging winds upon impinging on the surface.

*Corresponding author address: Kenneth L. Pryor, NOAA/NESDIS/STAR, 5200 Auth Rd., Camp Springs, MD 20746-4304; e-mail: Ken.Pryor@noaa.gov.

Large output values of each of the microburst index algorithms indicate that the ambient thermodynamic structure of the troposphere fits the prototypical environment for each respective microburst type (i.e. wet, hybrid, dry). Based on validation of the GOES WMSI and HMI products over the Oklahoma Panhandle during the 2005 and 2006 convective seasons, Pryor (2006b) noted a negative functional relationship between WMSI and HMI values for convective wind gusts of comparable magnitude. The statistically significant negative correlation between WMSI and HMI values likely reflects a continuum of favorable environments for downbursts, ranging from wet to intermediate (hybrid) (Caracena et al. 2007). The negative relationship between WMSI and HMI values associated with observed downbursts underscores the importance of the ambient thermodynamic structure of the boundary layer in the acceleration of convective downdrafts and resulting downburst wind gust magnitude. Numerous cases of hybrid microbursts have been documented (Ellrod 1989; Pryor 2006a; Pryor2006b). Among the most famous of this type of microburst studied was the Dallas-Fort Worth Microburst of August 1985 (Ellrod 1989). The hybrid microburst has been found to be a common type of microburst over the central and western United States, and thus necessitates the development of an algorithm to assess wind gust potential associated with this type of downburst environment.

Accordingly, a new diagnostic nowcasting product, the Microburst Windspeed Potential Index (MWPI), is derived from merging the WMSI and HMI algorithms. The MWPI is designed to quantify the most relevant factors in convective downburst generation in intermediate thermodynamic environments by incorporating CAPE, the sub-cloud lapse rate between the 670 and 850 mb levels, and the dew point depression difference between the typical level of a convective cloud base near 670 mb and the sub-cloud layer at 850 mb. In a prototypical dry microburst environment, Wakimoto (1985) identified a convective cloud base height near 500 mb associated with an “inverted V” profile. In contrast, Atkins and Wakimoto (1991) identified a typical cloud base height in a pure wet microburst environment near 850 mb. Thus, a cloud base height of 670 mb was selected for a hypothetical, weak shear intermediate microburst environment. This selection agrees well with the mean level of free convection (LFC) of 670 mb inferred from the inspection of twenty GOES proximity soundings corresponding to downburst events that occurred in Oklahoma between 1 June and 31 July 2005. In addition, Schafer (1986) noted that along the western periphery of the dryline in an environment with westerly geostrophic flow (favorable background flow in the dryline environment), the top of the mixed layer is typically found near 670 mb. In a free convective regime (i.e. light winds, no convective inhibition (CIN)), the mean LFC of 670 mb can be considered representative of convective cloud base heights in an intermediate microburst environment such that LFC ≈ LCL ≈ $z_i$ ≈ 670 mb, where LCL is the lifting condensation level and $z_i$ represents mixed layer depth. Johns and Doswell (1992) identified necessary ingredients for deep convection:

1. A moist layer of sufficient depth in the low or mid troposphere

2. A steep lapse rate to allow for a substantial "positive area" or CAPE

CAPE has an important role in precipitation formation due to the strong dependence of updraft strength and resultant storm precipitation content on positive buoyant energy. The subsequent precipitation loading initiates the convective downdraft which is sustained by sub-cloud evaporational cooling and negative buoyancy that accelerate the downdraft in the unsaturated layer. Collectively, evaporative cooling and resulting downdraft strength are enhanced by large liquid water content and related water surface available for evaporation, and a steep lapse rate that acts to maintain negative buoyancy as a downdraft descends in the sub-cloud layer. Ellrod (1989) noted that lapse rates, specifically between 700 and 850 mb, contained the most predictive information for determining downburst potential using GOES sounder profile data. Thus, the Microburst Windspeed Potential Index (MWPI) that accounts for both updraft (**U**) and downdraft instability (**D**) in microburst generation is defined as

$$MWPI \equiv \underbrace{\{(CAPE/100)\}}_{\mathbf{U}}+\underbrace{\{\Gamma+ (T-T_d)_{850}-(T-T_d)_{670}\}}_{\mathbf{D}}$$

where Γ is the lapse rate in degrees Celsius (C) per kilometer from the 850 to the 670 mb level, and the quantity $(T-T_d)$ is the dewpoint depression (C). The prototype sounding profile that illustrates the MWPI algorithm is displayed in Figure 1. Derivation of the MWPI algorithm is primarily based on parameter evaluation and pattern recognition techniques as employed in the severe convective storm forecasting process (Johns and Doswell 1992). Climatology of severe storm environmental parameters (Nair et al. 2002) has found that a deeper convective mixed layer, as represented by large LFCs, predominates in the warm season over the southern Plains. In fact, it was found by Nair et al. (2002) that moderately high LFCs that coexist with large CAPE over the Great Plains are associated with an observed maximum in severe convective storm occurrence.

Thus, this paper provides an initial assessment of the Microburst Windspeed Potential Index (MWPI), presents case studies demonstrating effective operational use of the MWPI product, and presents validation results for the 2007 convective season. Although there is not currently an observational requirement for microburst potential for the GOES-R Advanced Baseline Imager (ABI) (Schmit et al. 2005), the ABI does have promising capability to generate a sounding profile with greatly improved temporal and spatial resolution as compared to the existing GOES (8-P) sounders. In light of this capability, the eventual implementation of a sounder-derived microburst potential algorithm is feasible. Considering that seven of the sixteen bands of the ABI are in common with the bands of the heritage sounder, the ABI should effectively produce a sounding profile comparable in quality to the current GOES. The increase in temporal resolution should greatly aid the mesoscale forecaster (Johns and Doswell 1992) in the analysis of trends in thermodynamic environments.

## 2. METHODOLOGY

The objective of this validation effort was to qualitatively and quantitatively assess and intercompare the performance of the GOES sounder-derived microburst products by employing classical statistical analysis of real-time data. Accordingly, this effort entailed a study of downburst events over the southern High Plains during the summer of 2007 that was executed in a manner that emulates historic field projects such as the 1982 Joint Airport Weather Studies (JAWS) (Wakimoto 1985) and the 1986 Microburst and Severe Thunderstorm (MIST) project (Atkins and Wakimoto 1991). Data from the GOES HMI, MWPI and WMSI products was collected over the Oklahoma Panhandle and western Texas for downburst events that occurred between 1 June and 30 September 2007 and validated against surface observations of convective wind gusts as recorded by Oklahoma and West Texas Mesonet stations. Wakimoto (1985) and Atkins and Wakimoto (1991) discussed the effectiveness of using mesonet surface observations and radar reflectivity data in the verification of the occurrence of downbursts. Well-defined peaks in wind speed as well as significant temperature decreases (Wakimoto 1985; Atkins and Wakimoto 1991) were effective indicators of high-reflectivity downburst occurrence. As illustrated in the flowchart in Figure 11, images were generated in Man computer Interactive Data Access System (McIDAS) by a program that reads and processes GOES sounder data, calculates and collates microburst risk values, and overlays risk values on GOES imagery. Output images were then archived via FTP (ftp://ftp.orbit.nesdis.noaa.gov/pub/smcd/opdb/wmsihmiok/). Cloud-to-ground (CG) lightning data from the National Lightning Detection Network (NLDN) was plotted over GOES imagery to compare spatial patterns of CG lightning to surface observations of downburst wind gusts as described in Pryor (2006b).

Next Generation Radar (NEXRAD) base reflectivity imagery (levels II and III) from National Climatic Data Center (NCDC) was utilized to verify that observed wind gusts were associated with high-reflectivity downbursts and not associated with other types of convective wind phenomena (i.e. gust fronts). In addition, echo tops (ET) (level III) data were collected and analyzed to infer convective storm updraft intensity, based on the premise that higher echo tops are associated with more intense convective updrafts. NEXRAD images were generated by the NCDC Java NEXRAD Viewer (Available online at http://www.ncdc.noaa.gov/oa/radar/jnx/index.html). Another application of the NEXRAD imagery was to infer microscale physical properties of downburst-producing convective storms. Particular radar reflectivity signatures, such as the rear-inflow notch (RIN)(Przybylinski 1995) and the spearhead echo (Fujita and Byers 1977), were effective indicators of the occurrence of downbursts.

Since surface data quality is paramount in an effective validation program, the Oklahoma Panhandle and western Texas was chosen as a study region due to the wealth of high quality surface observation data provided by the Oklahoma and West Texas Mesonets (Brock et al. 1995; Schroeder et al. 2005), a thermodynamic environment typical of the High Plains region during the warm season, and relatively homogeneous topography. In addition, the dryline (Schafer 1986) is frequently observed over western Texas and the Oklahoma Panhandle during the warm season. Schafer (1986) noted that the dryline plays a significant role in convective storm climatology in the Southern Plains region in which convective storm activity developed on 70 % of dryline days. The High Plains region encompasses the Oklahoma Panhandle and western Texas that extends from 100 to 103 degrees West (W) longitude and is characterized by short-grass prairie. The treeless, low-relief topography that dominates the sparsely populated High Plains allows for the assumption of horizontal homogeneity when deriving a conceptual model of a boundary layer thermodynamic structure favorable for downbursts. More importantly, the topography over the Oklahoma-Texas Panhandle region facilitates relatively smooth flow (due to small surface roughness) with respect to downburst winds in which drag and turbulent eddy circulation resulting from surface obstructions (i.e., buildings, hills, trees) are minimized. Equation 4.6 in Sorbjan (1989) describes the wind profile in the surface layer and dictates that the relatively small roughness parameter of short grass prairie would permit wind gust measurements that are more representative of downburst intensity. Especially noteworthy in mesonet station siting is the superior wind exposure that satisfies World Meteorological Organization (1983) standards and minimal slope in the vicinity of each station that reduces topographic effects that may influence downburst magnitude and resulting measured surface wind gusts. Shafer et al. (1993) addressed the preference of rural station siting to avoid anthropogenic factors, especially factors that may influence convective wind gust magnitude as recorded by mesonet stations. Site characteristics, data quality assurance, and wind sensor calibration are thoroughly documented in Brock et al. (1995) and Schroeder et al. (2005).

Downburst wind gusts, as recorded by mesonet observation stations, were measured at a height of 10 meters (33 feet) above ground level. In order to assess the predictive value of GOES microburst products, the closest representative index values used in validation were obtained for retrieval times one to three hours prior to the observed surface wind gusts. Representativeness of proximate index values was ensured by determining from analysis of surface observations, and radar and satellite imagery, that no change in environmental static stability and air mass characteristics between product valid time and time of observed downbursts had occurred. Furthermore, in order for the downburst observation to be included in the validation data set, it was required that measurable precipitation be observed by each respective mesonet station and the parent convective storm cell of each downburst, with radar reflectivity greater than 35 dBZ, be located overhead at the time of downburst occurrence. An additional criterion for inclusion into the data set was a wind gust measurement of at least F0 intensity (35 knots) on the Fujita scale (Fujita 1971). Wind gusts of 35 knots or greater are considered to be operationally significant for transportation, especially boating and aviation. An algorithm devised by Wakimoto (1985) to visually inspect wind speed observations over the time intervals encompassing candidate downburst events was implemented to exclude gust front events from the validation data set. Since nocturnal inversion strength typically maximizes between 0300 and 0700 CDT (0800 and 1200 UTC) during the convective season, as inferred from inspection of West Texas Mesonet observations, it was decided to include only those downburst events that occurred during afternoon daylight hours (1700 to 0300 UTC). This requirement was designed to eliminate the influence of nocturnal temperature inversions and resulting static stability in the boundary layer on downburst wind gusts observed at the surface. In summary, the screening process employed to build the validation data set that consisted of criteria based on surface weather observations and radar reflectivity data yielded a sample size of 35 downbursts and associated index values.

At this early stage in the algorithm assessment process, it is important to consider covariance between the variables of interest: MWPI and surface downburst wind gust speed. A very effective means to assess the quantitative functional relationship between microburst index algorithm output and downburst wind gust strength at the surface is to calculate correlation between these variables. Thus, correlation between GOES HMI, WMSI, and MWPI values and observed surface wind gust velocities for the selected 35 events were computed and intercompared to assess the significance these functional relationships. Statistical significance testing was conducted, in the manner described in Pryor and Ellrod (2004), to determine the confidence level of correlations between observed downburst wind gust magnitude and microburst risk values. Hence, the confidence level is intended to quantify the robustness of the correlation between microburst index values and wind gust magnitude.

## 3. CASE STUDIES

### 3.1 *Oklahoma Panhandle Downbursts*

During the afternoon of 26 July and 23 August 2007, convective storms developed along a cold front that extended from southeastern Colorado into northern New Mexico. Particularly intense storms tracked east into the western Oklahoma Panhandle, producing strong downburst wind gusts of 49 knots on 26 July and 41 knots on 23 August as observed by the Kenton, Oklahoma Mesonet station. Comparison of late afternoon GOES MWPI imagery to the location of the observed downbursts revealed that the MWPI effectively indicated the potential for strong downburst wind gusts.

Apparent in Figure 1, the MWPI product image at 2200 UTC 26 July 2007, is a line of convective storms developing along a cold front that extends from southeastern Colorado to central New Mexico. In addition, an MWPI value of 54 near Kenton was the highest indicated over the Oklahoma Panhandle at 2200 UTC, about two hours prior to downburst occurrence. Based on linear regression, the MWPI value of 54, plotted in orange, indicated the potential for severe downburst wind gusts of 50 knots or greater. Color-coded markers in the microburst product images represent cloud-to-ground (CG) lightning discharges with the following progression during the hour succeeding product valid time: first ten-minute period (:00 to :10), white; second ten-minute period (:10 to :20), blue; third ten-minute period (:20 to :30), red; fourth ten-minute period (:30 to :40), green; fifth ten-minute period (:40 to :50), yellow; sixth ten-minute period (:50 to :00), cyan.

By 0000 UTC 27 July, a convective storm propagated eastward into the western Oklahoma Panhandle, producing a downburst wind gust of near 50 knots at Kenton mesonet station. Amarillo, Texas (KAMA) NEXRAD reflectivity image at 0005 UTC, the time of downburst occurrence, displayed in Figure 2 a high-reflectivity storm echo (> 50 dBZ) over Kenton mesonet station. Co-located with the high radar reflectivity over the station was a maximum echo top exceeding 40000 feet. Also apparent in the MWPI image is a low cloud-to-ground (CG) lightning flash density associated with the downburst-producing convective storm near Kenton. Between 0000 and 0010 UTC, only three CG discharges (one positive, two negative, white markers) were recorded in close proximity to the convective storm, as shown in Figure 3. Pryor (2006b) described the role of an elevated charge dipole in the suppression of CG lightning within downburst-producing convective storms. An elevated dipole resulting from intense updrafts most likely effected suppressed CG lightning activity associated with the downburst observed at Kenton as evidenced by storm radar echo top heights near 45000 feet, well in excess of the height of the -20C isotherm at 20715 feet as indicated in Figure 1, a GOES sounding over Clayton, New Mexico at 2200 UTC. Pryor (2006b) noted a height difference between the -20C isotherm and the storm echo top greater than 25000 feet associated with an elevated dipole. The comparison of echo top heights to CG lightning flash rates in downburst-producing convective storms emulates the study of an Alabama hailstorm (Williams 2001). Williams (2001) found that suppressed CG flash rates in the hailstorm were associated with echo tops heights that approached 45000 feet. The sole positive discharge near downburst occurrence was likely the result of one of two possible mechanisms as outlined in Williams (2001): local unshielding of upper positive charge, or misidentification of an intracloud flash by the NLDN. The precipitation unshielding hypothesis seeks to account for the suppression of negative CG flashes in heavy precipitation-producing storms and may present a more pertinent explanation for the observed CG flash rate associated with the 26 July downburst.

In a similar manner to the 26 July downburst, the 23 August downburst at Kenton mesonet station occurred during the early evening after a maximum in surface heating and near a local maximum in MWPI values (43) as shown in Figure 4. At the time of peak wind (41 knots, 2302 UTC), radar reflectivity imagery in Figure 5 from Amarillo NEXRAD displayed a distinctive bow echo (Przybylinski 1995) over the Kenton mesonet station with a trailing RIN, indicating that the peak wind gust recorded was associated with a convective downburst. Also similar to the 26 July event, storm echo top heights approached 45000 feet at the time of downburst occurrence, well in excess of the height of -20C isotherm near 21000 feet as identified in Figure 4, a GOES sounding from nearby Clayton, New Mexico at 2100 UTC. Since no CG lightning was observed by the NLDN near the time of the downburst wind gust in the vicinity of Kenton mesonet station, it could be inferred again that an elevated electric charge dipole was coincident with downburst occurrence.

Inspection of late afternoon GOES sounding profiles for both events over Clayton revealed the presence of a well-developed mixed layer and a large CAPE-inverted V profile. Common to both events, the sounding profiles, similar to the type A (Wakimoto 1985), reflected the favorability for downbursts in an intermediate thermodynamic environment as indicated in GOES MWPI imagery two hours prior. The surface dewpoint depression, as observed by the Kenton mesonet station and shown in Figure 6, echoed the development and evolution of the mixed layer during the afternoon hours with a dewpoint depression of 40F indicated by 2100 UTC. The Oklahoma Panhandle downburst events reiterated the importance of storm precipitation content that was related to large CAPE and resulting updraft strength. Also important was the boundary layer structure that fostered evaporational cooling and resultant negative buoyancy as precipitation descended in the sub-cloud layer. The positive correlation between index values and wind gust magnitude for these events reflected the effectiveness of the MWPI product in the short-term forecasting of potential downburst magnitude.

### 3.2 *West Texas Dryline Downbursts*

Strong downbursts were observed over western Texas during the afternoon of 6 September 2007. During the afternoon of 6 September, convective storms developed near the dryline over western Texas. Strong downburst wind gusts, recorded by Slaton, Clarendon, and McLean (West Texas) mesonet stations, were observed in close proximity to local maxima in GOES MWPI values, indicated in product imagery one to two hours prior to the observed events. These downburst events demonstrated strong correlation between MWPI values and the magnitude of convective wind gusts recorded by West Texas Mesonet stations.

The dryline was especially apparent in HMI and MWPI imagery during the afternoon of 6 September 2007 as a zone of enhanced cumulus, resulting from mixing and convergence, and developing convective storm activity extending from north to south over the Texas Panhandle. Figure 7, the 2000 UTC MWPI image, displayed a regional maximum in index values along the dryline over western Texas reflecting an increase in vertical mixing and a resultant increase in CBL depth. Figure 8, an analysis of surface observations from the Oklahoma Mesonet confirmed the presence of the dryline over the panhandle region as a zonal gradient in dewpoint temperatures. The importance of the dryline in initiating convective storm activity and fostering a favorable environment for downbursts was discussed earlier in this paper. Ziegler and Hane (1993), in their observational study of a western Oklahoma dryline, found that the environment is dominated by vertical mixing that maintains a convective boundary layer (CBL) on both flanks of the dryline. The dryline resembles a “mixing zone” that slopes eastward from the surface dryline location, then becomes a quasi-horizontal elevated moist layer above the CBL east of the dryline. The elevated MWPI values east of the dryline in the vicinity of downburst occurrence serve as evidence of the presence of a well-developed mixed layer and a favorable environment for downbursts resulting from sub-cloud evaporation of precipitation.

Between 2000 and 2200 UTC 6 September 2007, the area of convective storms propagated eastward across the panhandle and south plains regions. An isolated and particularly intense convective storm produced the first severe downburst of the afternoon as observed by the Slaton mesonet station, in southeastern Lubbock County, at 2040 UTC. The 2000 UTC MWPI product image in Figure 7 indicated a proximate value of 48, considered most representative of the undisturbed environment in the vicinity of downburst occurrence. Based on linear regression, as will be discussed in the next section, an MWPI value of 48 is correlated with downburst wind gust potential of 48 knots, very close to the observed wind gust of 52 knots at Slaton. Radar reflectivity imagery from Lubbock, Texas (KLBB) NEXRAD confirmed in Figure 9 that the recorded wind gust was associated with a convective storm. A high-reflectivity spearhead echo (Fujita and Byers 1977) was apparent in reflectivity imagery over Slaton at 2038 UTC, approximately two minutes before the downburst was observed at the surface. In addition, inspection of observations from Slaton mesonet station revealed a well-defined peak in wind speed coincident with a peak in rainfall rate at the time of downburst occurrence:

Slaton 2NE Data from 09/06/2007

| | | TEMP | | ---WIND------ | | |
|---|---|---|---|---|---|---|
| TIME | RAIN? | (F) | DEW | DIR | SPD | PK |
| (UTC) | | 6ft | POINT | (MPH) | | |
| 2045 | RAIN | 70 | 66 | 063 | 9 | 21 |
| 2040 | RAIN | 75 | 66 | 178 | 30 | 60 |
| 2035 | RAIN | 87 | 69 | 181 | 19 | 36 |

By 2100 UTC, as shown in Figure 7, convective storms over the Texas Panhandle were merging into a cluster east of Amarillo. In addition, MWPI values were increasing to greater than 60 downstream of the convective storm cluster, indicating the high likelihood for severe downbursts. A stronger downburst wind gust of 58 knots was observed at McLean at 2220 UTC. Similar to the previously observed downbursts, radar reflectivity imagery from Amarillo, Texas (KAMA) NEXRAD in Figure 9 displayed a spearhead echo and trailing RIN in the vicinity of McLean near the time of downburst occurrence. Observations from Clarendon and McLean also indicated sharp peaks in wind speed and rainfall rates coincident with observed downbursts:

Clarendon 2WSW Data from 09/06/2007

| | | TEMP | | ---WIND------ | | |
|---|---|---|---|---|---|---|
| TIME | RAIN? | (F) | DEW | DIR | SPD | PK |
| (UTC) | | 6ft | POINT | (MPH) | | |
| 2140 | RAIN | 72 | 64 | 204 | 32 | 45 |
| 2135 | RAIN | 76 | 64 | 209 | 45 | 57 |
| 2130 | RAIN | 86 | 67 | 192 | 19 | 39 |

McLean 1E Data from 09/06/2007

| | | TEMP | | ---WIND------ | | |
|---|---|---|---|---|---|---|
| TIME | RAIN? | (F) | DEW | DIR | SPD | PK |
| (UTC) | | 6ft | POINT | (MPH) | | |
| 2225 | RAIN | 65 | 64 | 174 | 35 | 48 |
| 2220 | RAIN | 65 | 64 | 200 | 55 | 67 |
| 2215 | RAIN | 68 | 64 | 218 | 46 | 63 |

These downburst events over western Texas demonstrated the importance of boundary layer structure in the generation of downbursts. GOES sounding profiles at Lubbock at 2000 UTC and at Amarillo at 2100 UTC 6 September 2007 (not shown) displayed prototype large CAPE-inverted V profiles with LFCs near 670 mb. Both downburst events occurred in environments characterized by strong static instability that favored high-reflectivity storms with heavy rainfall as indicated by MWPI values greater than 40. Downburst-producing storms had a large precipitation content that fostered evaporational cooling and resultant negative buoyancy as precipitation descended in the sub-cloud layer. Also, downbursts that were observed at Slaton and McLean were observed near local maxima in MWPI values.

## 4. DISCUSSION

Validation based on the qualifying 35 downburst events documented over the Oklahoma Panhandle and western Texas during the 2007 convective season indicated a stronger correlation ($r=.77$, $r^2=.60$) between MWPI values and observed surface downburst wind gusts than the correlation calculated between the WMSI and observed winds ($r=.65$). A scatterplot with an overlying regression line, featured in Figure 10, most effectively illustrates the strong positive relationship between MWPI values and wind gust magnitude, in which 60% of the variability in wind gust speed is coupled with variability in MWPI. The scatterplot was constructed by designating the X axis for the independent variable, MWPI, and the Y axis for the dependent variable, observed wind gust speed. In addition, hypothesis testing confirmed the statistical significance of this correlation with a 97% confidence level that the correlation represented a physical relationship between the MWPI and the strength of downburst wind gusts observed at the surface.

Eighteen representative proximity soundings were inspected to determine LFC heights for respective downburst events during the 2007 convective season. The histogram shown in Figure 10 that displays the number of soundings with respective LFC heights between 620 and 700 mb indicates a median and mode LFC of 670 mb. In addition, the mean LFC computed for the convective season was near 670 mb. Thus, in the majority of cases analyzed, the selection of 670 mb as the prototypical cloud base height over the southern High Plains was affirmed by LFCs observed between 620 and 700 mb as well as a symmetrical distribution in which the mean, median, and mode LFC was near 670 mb.

Diurnal and inter-diurnal variability in downburst activity over the Panhandle/West Texas domain during the 2007 convective season was also considered and analyzed. The inter-diurnal trend in downburst activity, illustrated in Figure 10, shows a multi-modal distribution with peaks in activity occurring in June, early July, late August and September. The period of relative inactivity that extended from mid-July through late August coincided with a persistent upper-level ridge over the southern Plains. The upper-level ridge with associated subsidence resulted in suppression of deep convective storm development within the domain. Any orographic convective storm activity that developed during this period was most often steered to the north of the study region. It is apparent that the late season increase in downburst frequency reflected an increase in frontal and dryline activity over the Panhandle/West Texas region as the persistent upper-level ridge gradually weakened during late August and September. In fact, the dryline was associated with 22 of the 35 (>60%) downburst events documented during the 2007 convective season. This finding highlighted the importance of the dryline as not only an initiator of convection over the High Plains, but also as a facilitator for downburst generation. As exemplified in the West Texas downburst case study in section 3.2, higher MWPI values were typically found in proximity to the dryline.

Analysis of diurnal trends in downburst activity revealed an expected late afternoon and early evening maximum. The diurnal variation in downburst activity, illustrated in the histogram in Figure 10, is consistent with the trend observed during the 1982 JAWS project and documented in Wakimoto (1985). In a similar manner to the observations examined in the JAWS study, the maximum in diurnal activity during the 2007 convective season over the Panhandle/West Texas mesonet domain occurred subsequently to a maximum in solar heating. This suggests that daytime solar heating of the boundary layer and the resulting thermodynamic profile is an important consideration in the assessment of downburst potential.

In addition to the primary factors identified in downburst generation over the High Plains during the 2007 convective season, including the presence of large CAPE and a deep, dry convective mixed layer, the presence of a mid-tropospheric layer of dry air near the 400 mb level was readily apparent in analyzed proximity sounding profiles. Previous studies of hybrid microbursts (Ellrod 1989; Pryor 2006a; Pryor 2006b) highlighted the importance of lateral entrainment of mid-tropospheric dry air into the precipitation core of the convective storm in downdraft acceleration. In this process, convective downdraft acceleration is the result of evaporational cooling and subsequent generation of negative buoyancy. The presence of mid-tropospheric dry air and its influence on downburst generation identified in this study may warrant the parameterization of an additional term to be included in the MWPI algorithm.

## 5. SUMMARY AND CONCLUSIONS

This validation effort and study of high-reflectivity downbursts over the southern High Plains reveals common environmental characteristics: significant CAPE, cloud base heights near 670 mb, and a dry-adiabatic, convective boundary layer. This favorable thermodynamic structure for downbursts is most effectively parameterized by the MWPI algorithm. As exemplified in the case studies and statistical analysis, the GOES MWPI product demonstrated utility in the short-term prediction of downburst magnitude.

The scatterplot diagram in Figure 10 illustrates that MWPI values do not represent absolute wind gust speeds, but rather quantitatively indicate relative convective wind gust potential. The statistical correlation approach allows the user to tune the relationship between MWPI values and downburst wind gust speeds according to local climatology. Thus, future research activity will entail real-time verification of the MWPI and WMSI products over diverse geographic regions that are covered by high-quality surface observation networks. Furthermore, the MWPI will be evaluated for application over the eastern United States. One important goal of this validation effort will be to derive a relationship between MWPI values and ordinal categories of microburst wind gust potential (<35 knots, 35-49 knots, 50-64 knots, >65 knots) in a similar manner to the process outlined for the WMSI product in Pryor and Ellrod (2004). As noted in the previous section, the importance of mid-tropospheric dry air entrainment in the downburst generation process may require the parameterization of an additional term in the MWPI algorithm. Overall, the MWPI product demonstrates a significant improvement over the WMSI product in assessment capability of potential downburst magnitude over the Great Plains. In addition to the improved spatial and temporal resolution that will be offered by GOES-R ABI soundings, the MWPI product shows promise to be an effective tool for downburst potential forecasting over the central and western United States.

**Acknowledgements**

The author thanks Mr. Derek Arndt (Oklahoma Climatological Survey)/Oklahoma Mesonet, and **Dr. John Schroeder** (Texas Tech University)/West Texas Mesonet for the surface weather observation data used in this research effort. The author also thanks Jaime Daniels (NESDIS) for providing GOES sounding retrievals displayed in this paper. Cloud-to-ground lightning data was available for this project through the courtesy of Vaisala, Inc.

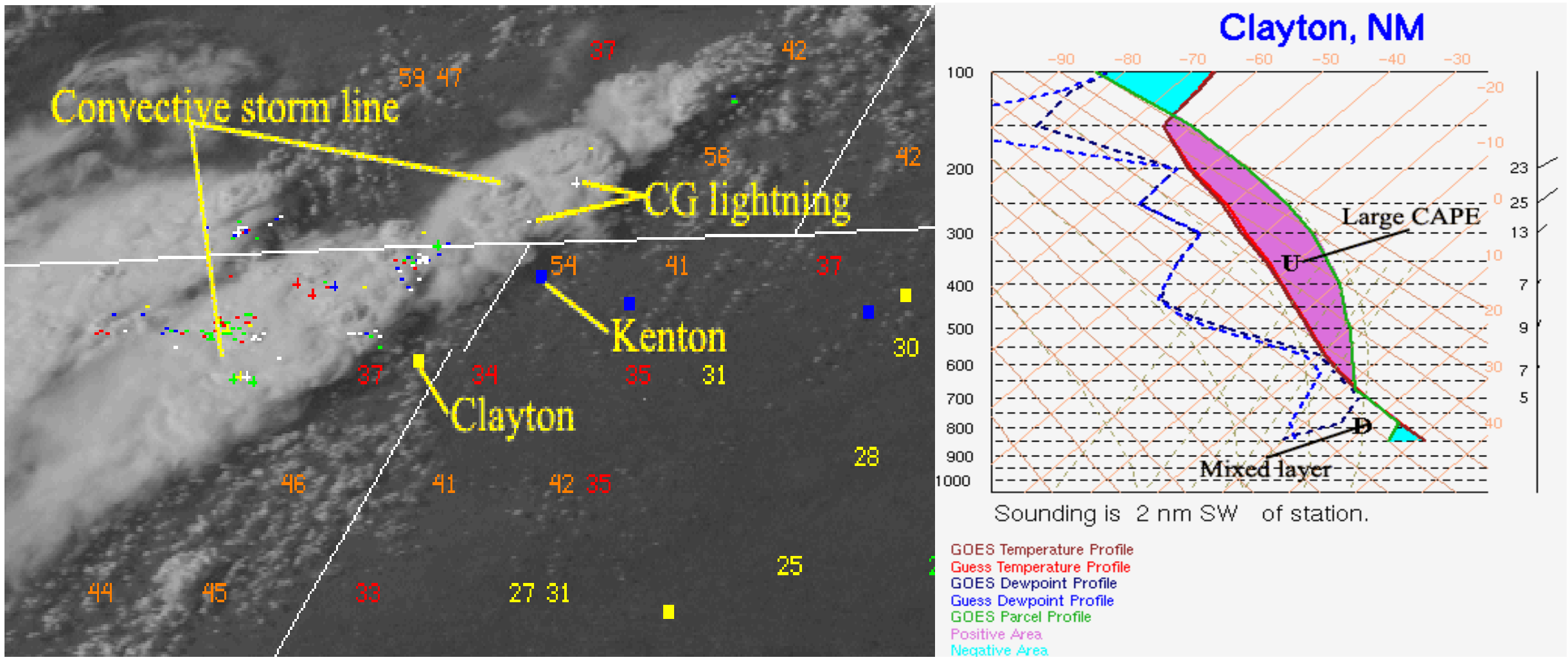

Figure 1. GOES MWPI product image at 2200 UTC 26 July 2007 (left) and corresponding prototypical GOES sounding profile at Clayton, New Mexico (right).

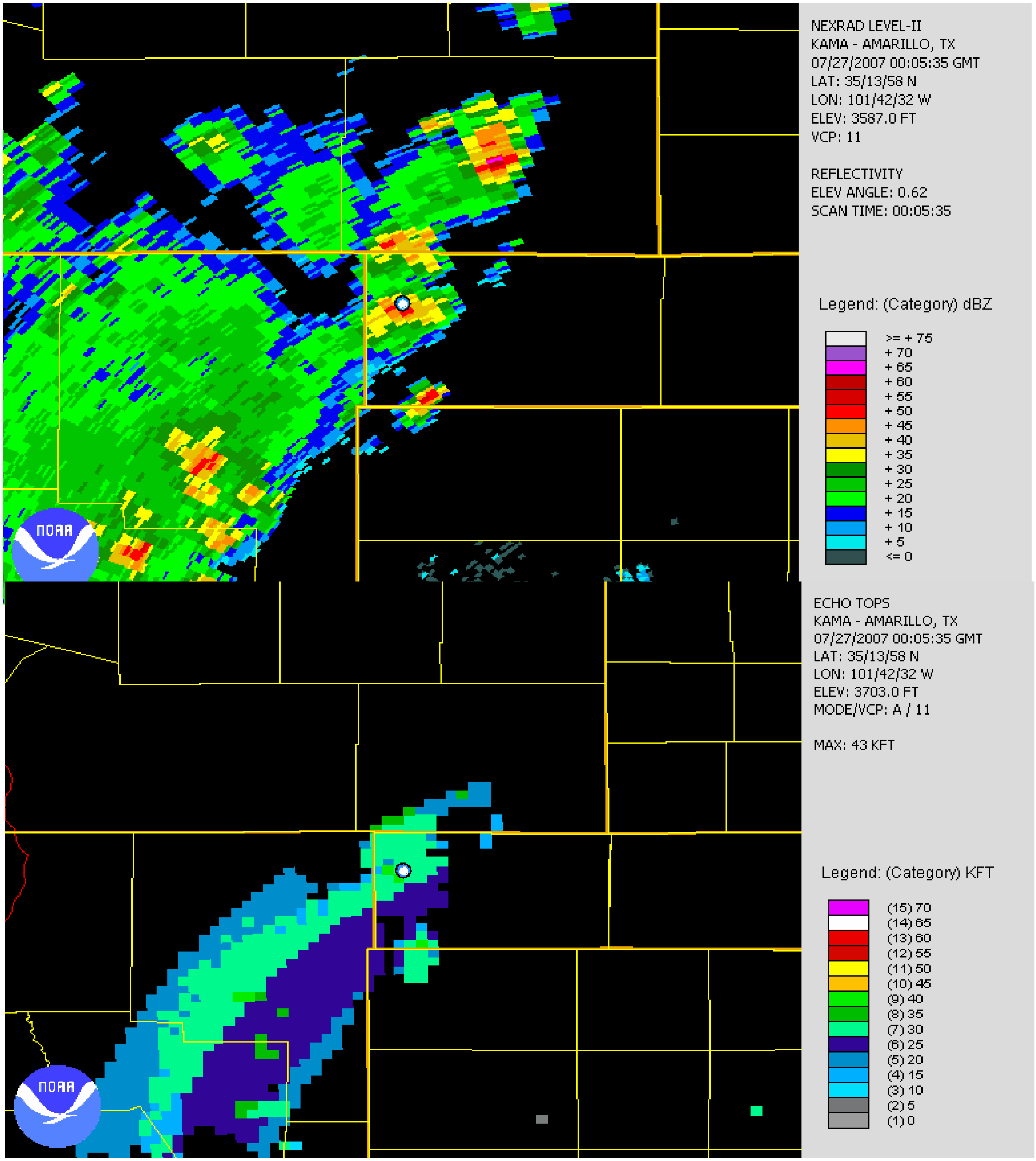

Figure 2. NEXRAD reflectivity imagery (top) and echo tops (bottom) at 0005 UTC 27 July 2007. Marker indicates the location of Kenton mesonet station.

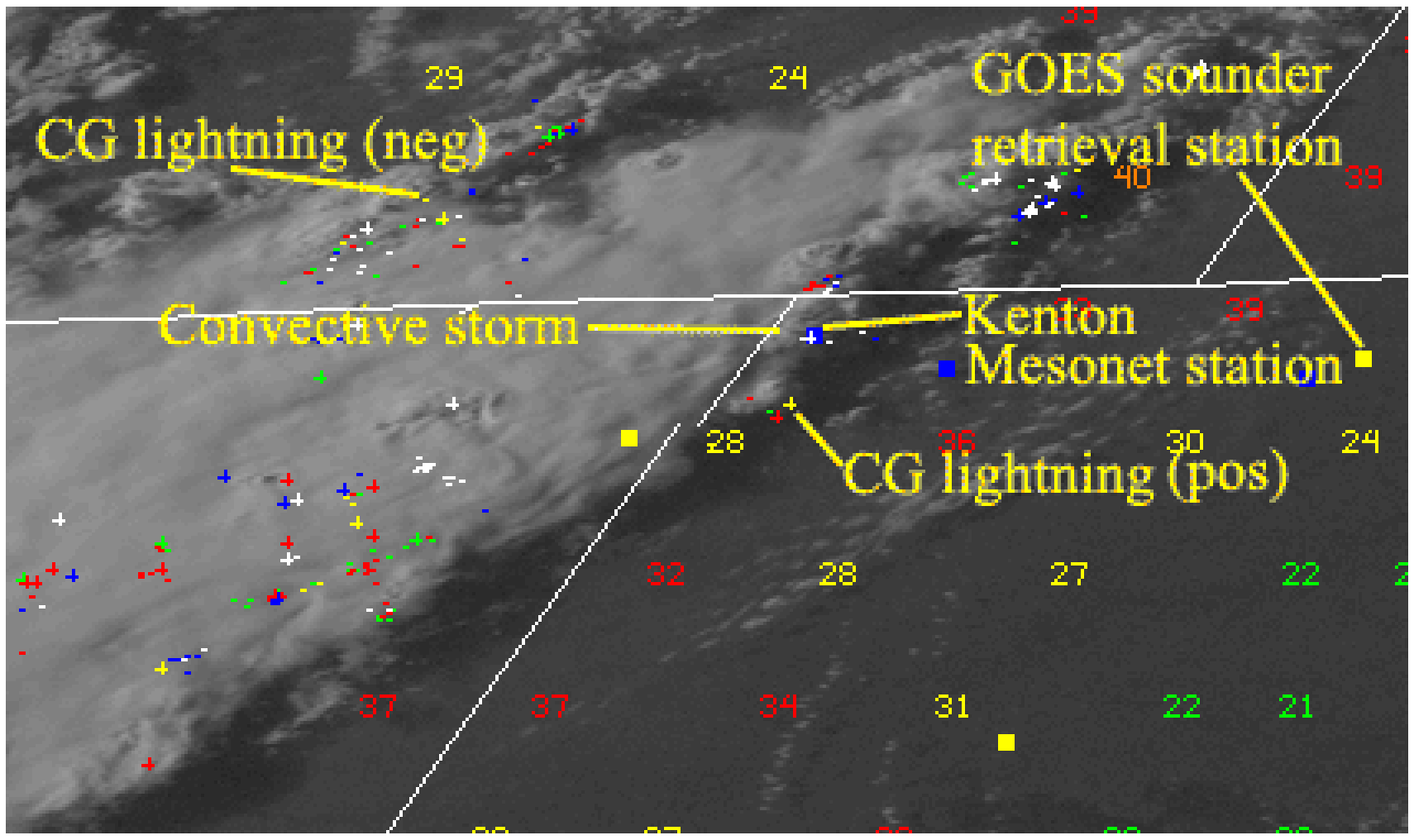

Figure 3. GOES MWPI product image at 0000 UTC 27 July 2007.

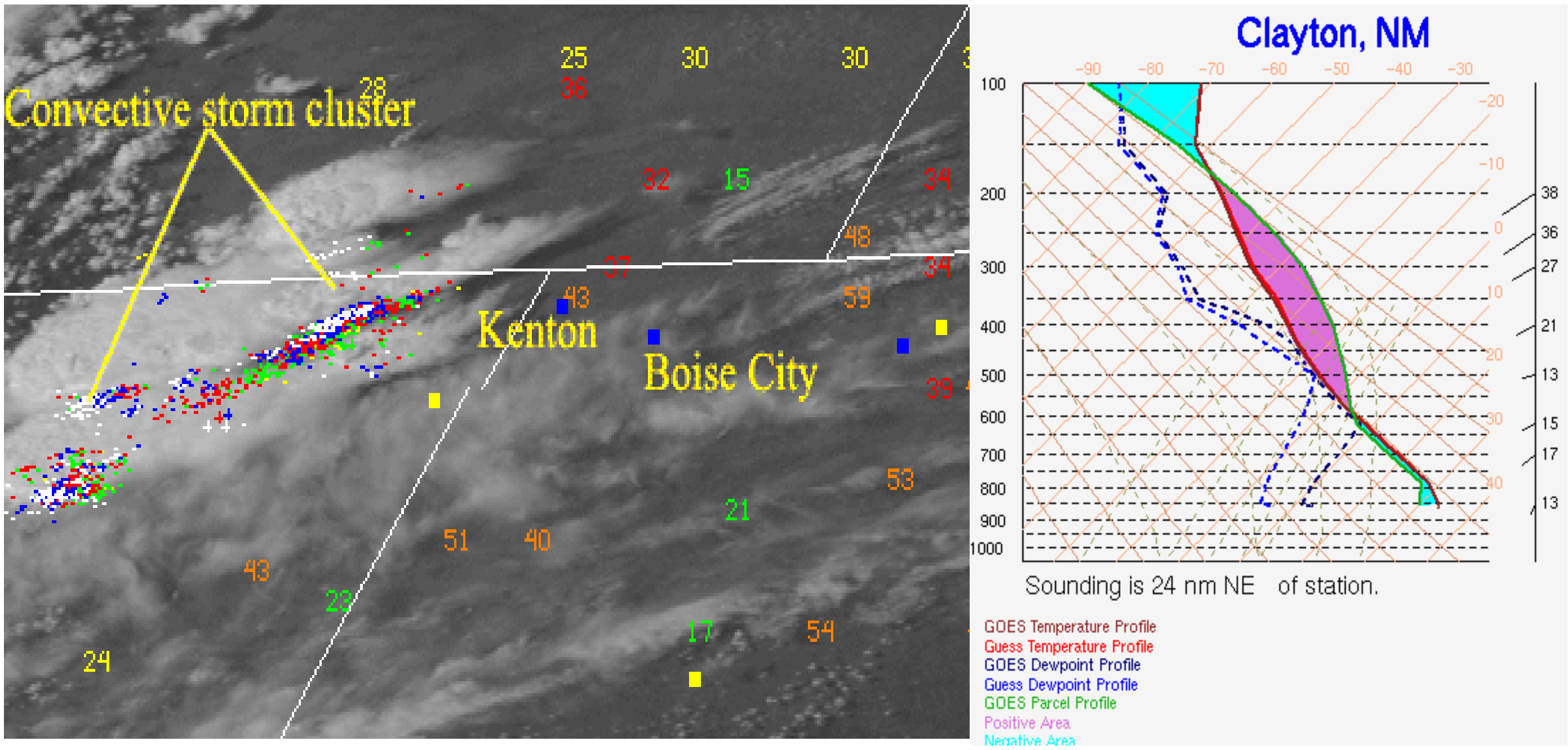

Figure 4. GOES MWPI product image at 2100 UTC 23 August 2007 (left) and corresponding GOES sounding profile at Clayton, New Mexico (right).

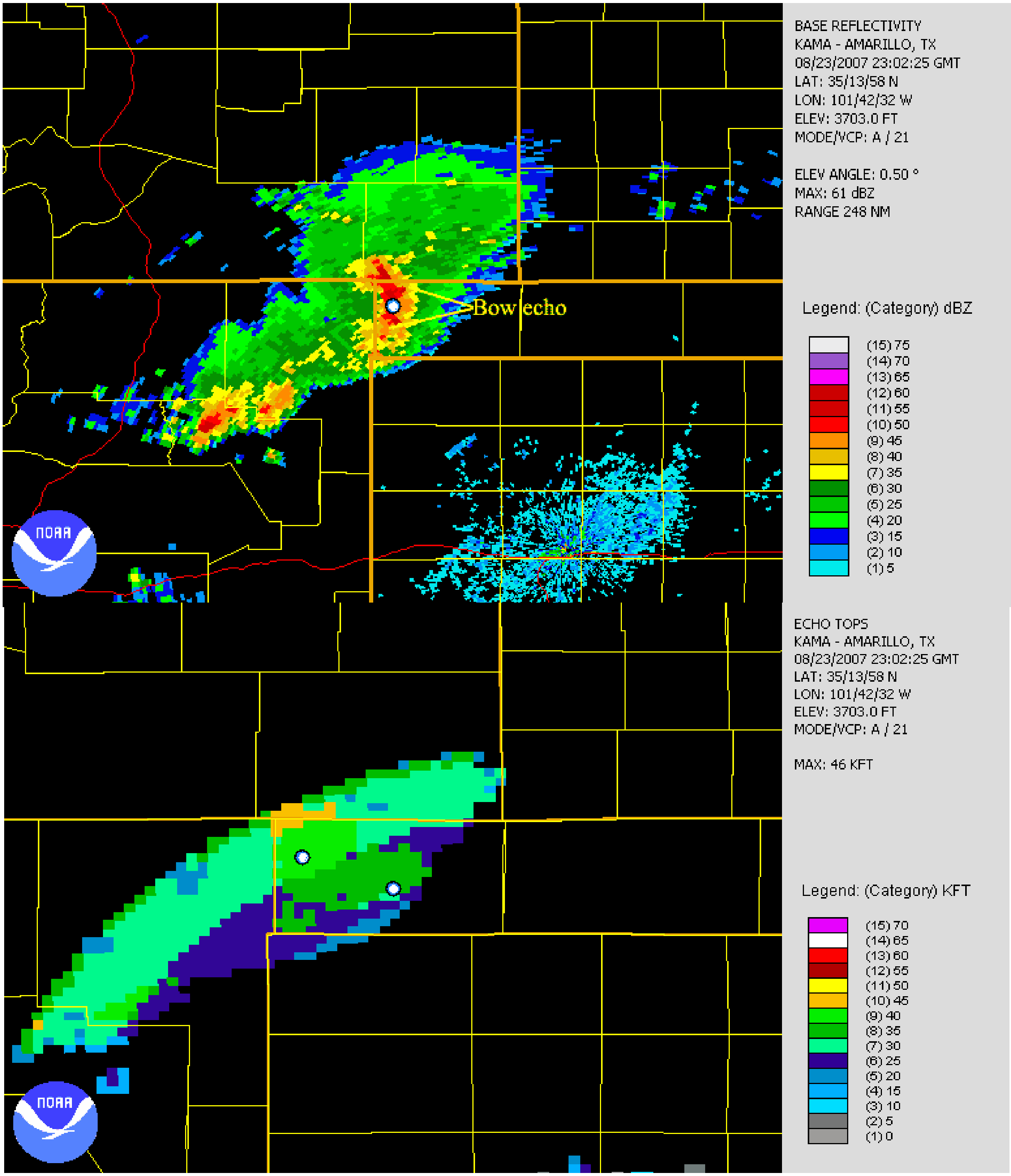

Figure 5. NEXRAD reflectivity imagery (top) and echo tops (bottom) at 2302 UTC 23 August 2007. Marker indicates the location of Kenton mesonet station.

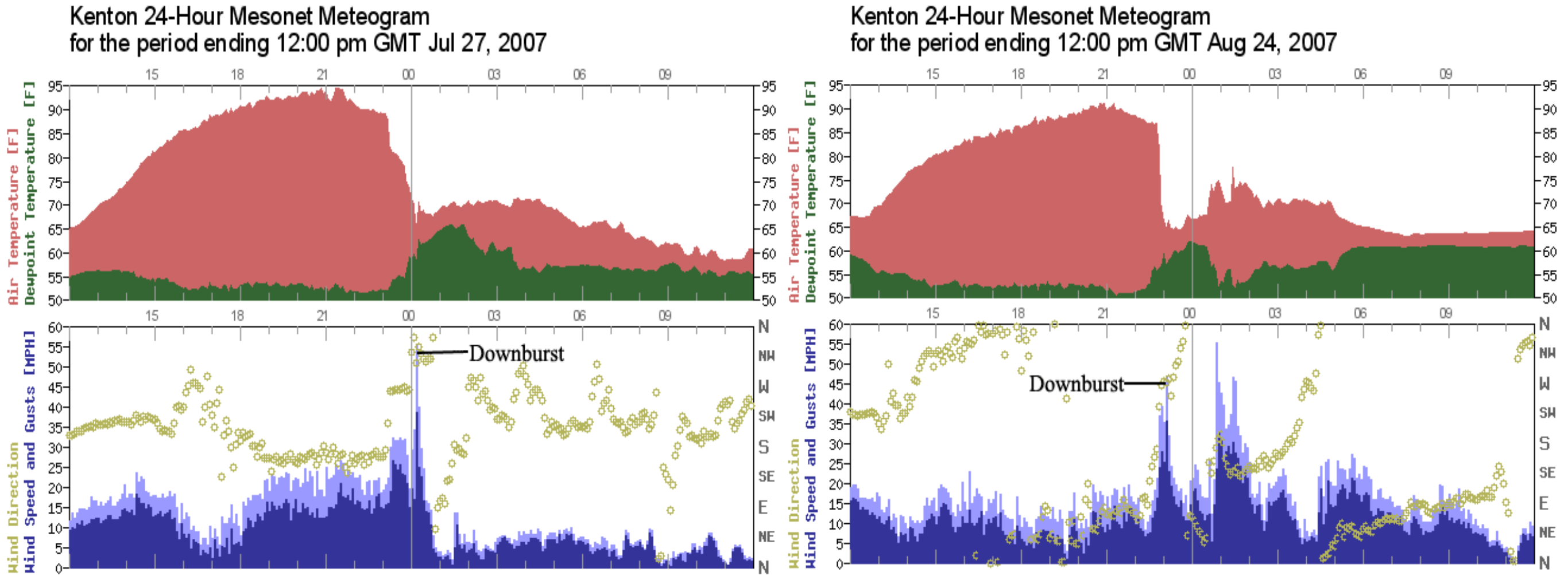

Figure 6. Oklahoma Mesonet meteograms from Kenton on 26 July (top) and 23 August 2007 (bottom).

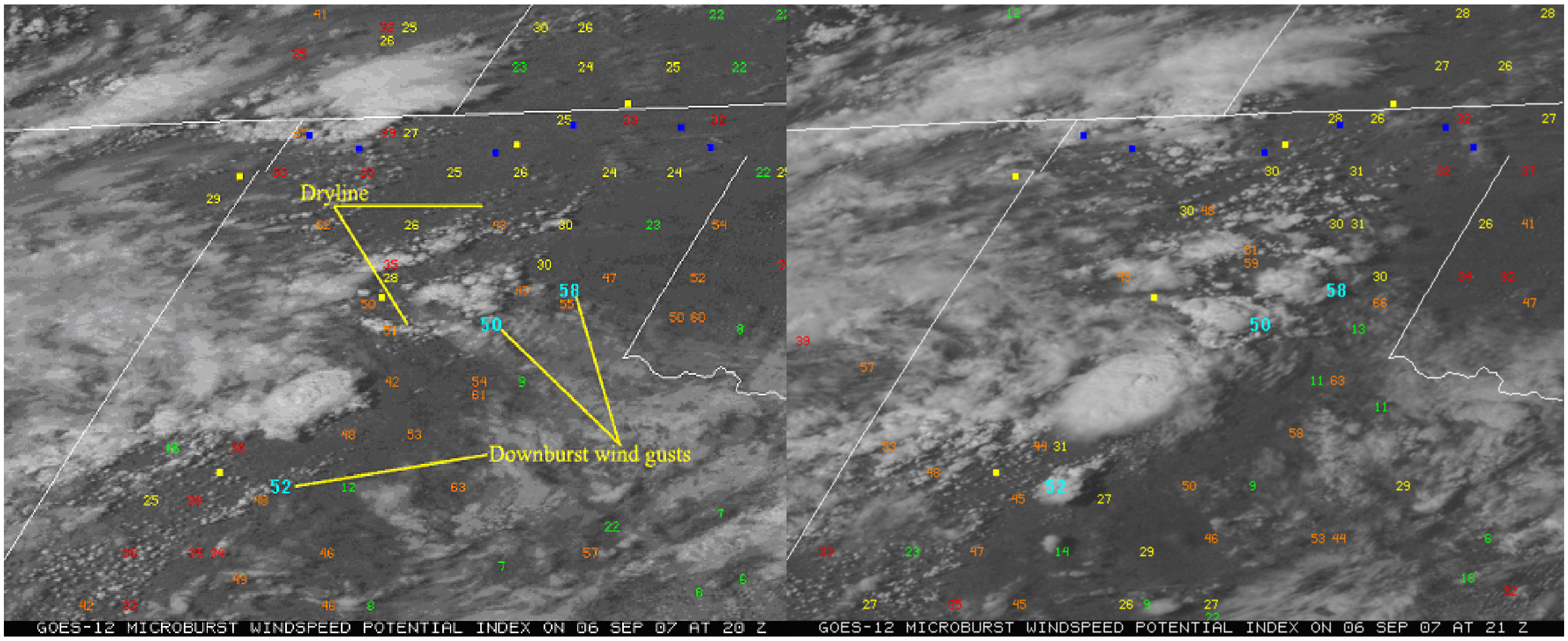

Figure 7. GOES MWPI product image at 2000 UTC (left) and 2100 UTC (right) 6 September 2007.

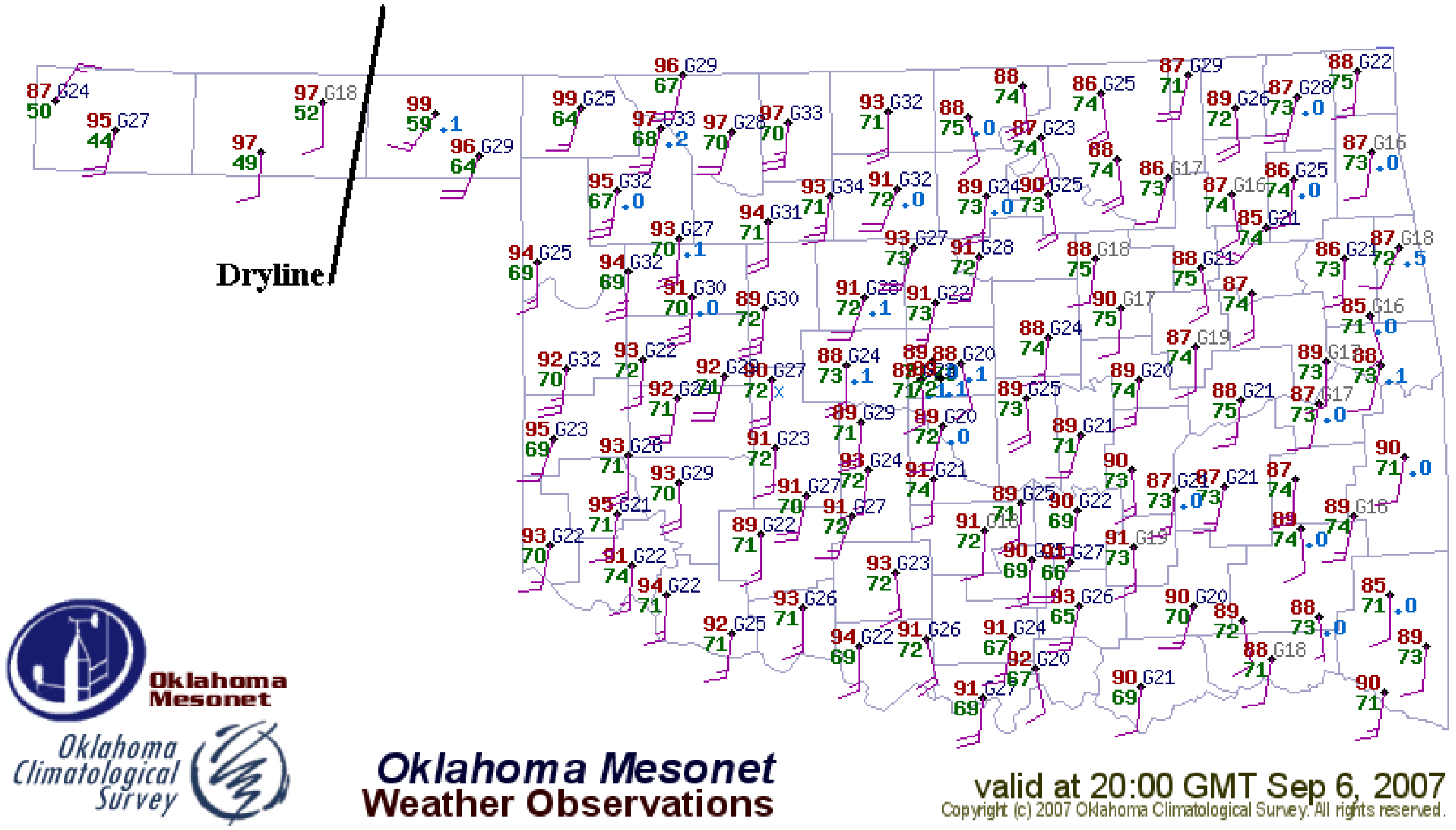

Figure 8. Surface analysis at 2000 UTC 6 September 2007 displaying the location of the dryline over the Oklahoma Panhandle.

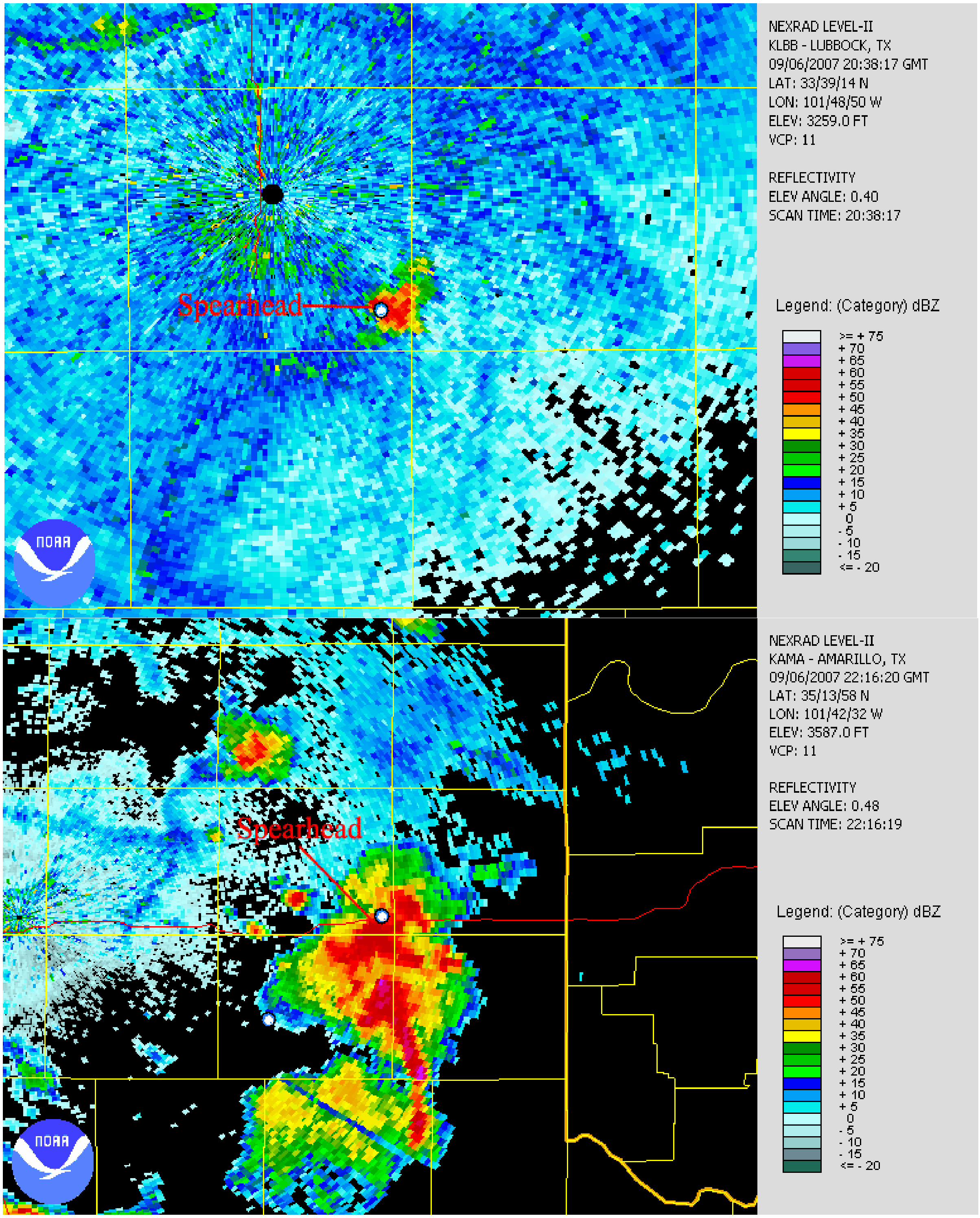

Figure 9. NEXRAD reflectivity imagery at 2038 UTC (top) and 2216 UTC (bottom) 6 September 2007. Markers indicate the location of Slaton and McLean mesonet stations.

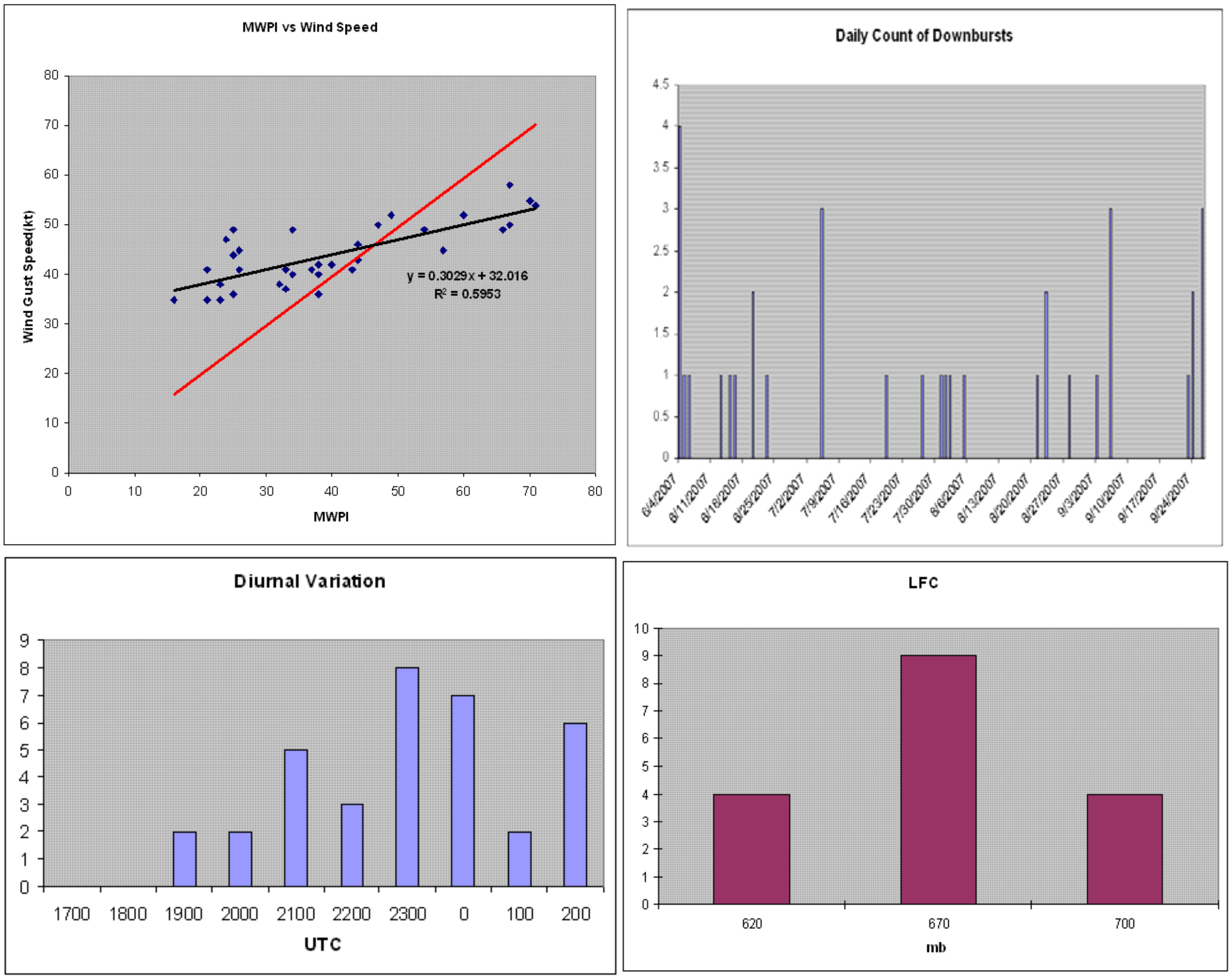

Figure 10. Statistical analysis of validation data over the Oklahoma Panhandle and western Texas domain between June and September 2007: Scatterplot of MWPI values vs. measured convective wind gusts for 35 downburst events (upper left), daily count of verified downbursts (upper right), diurnal variation of verified downbursts (lower left), and count of observed LFC heights from 17 GOES proximity soundings. The red line in the scatterplot indicates a perfect positive correlation (r=1.0).

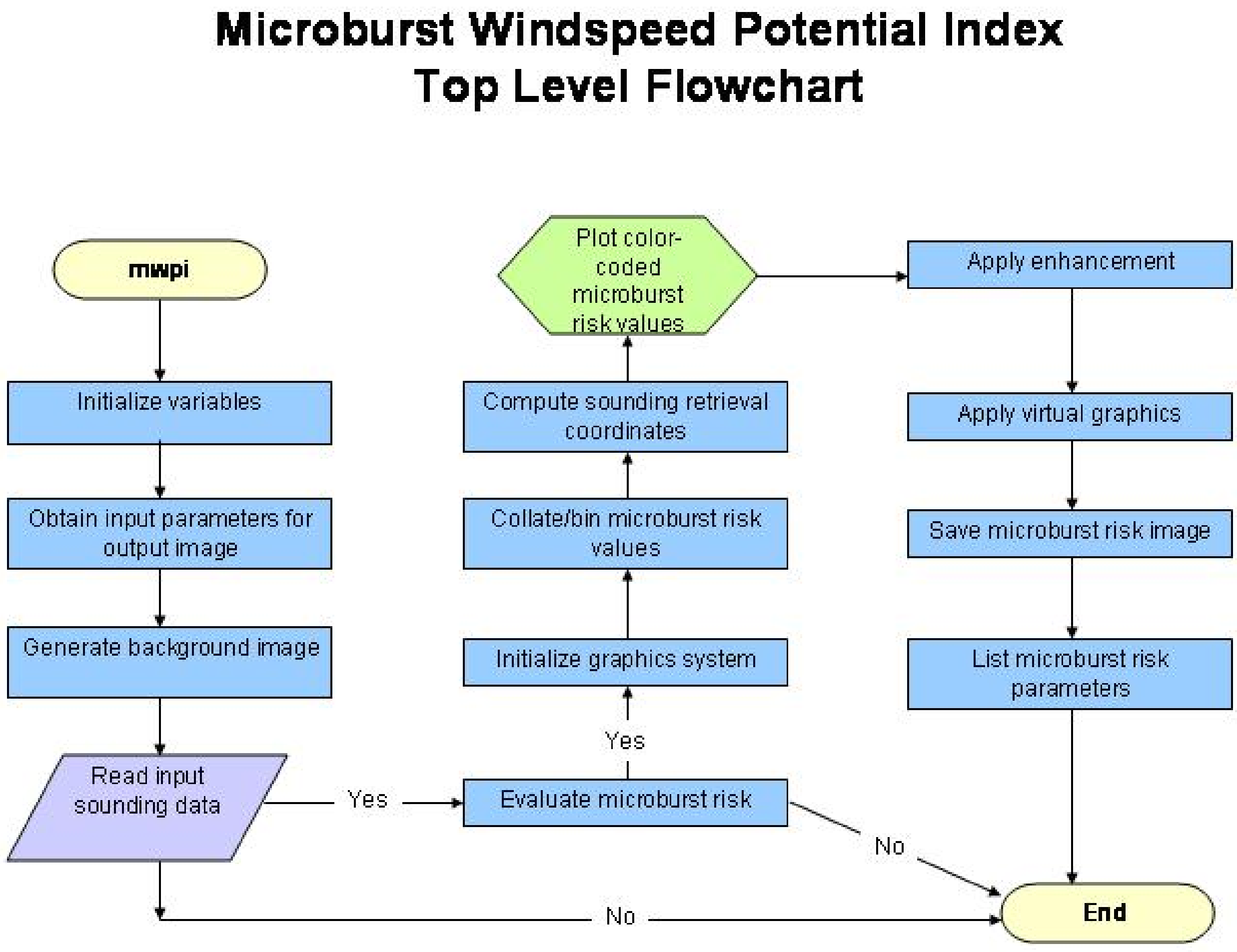

Figure 11. Flowchart illustrating the operation of the MWPI program.